\documentclass[namedreferences]{solarphysics}

\usepackage[hyperref,optionalrh,showbiblabels]{spr-sola-addons} % For Solar Physics 
\usepackage{graphicx}        % For eps figures, newer & more powerfull
\usepackage[T1]{fontenc}
\usepackage{color}           % For color text: \color command
\usepackage{breakurl}        % For breaking URLs easily trough lines
\usepackage{hyperref}

\usepackage{hyperref} 
\hypersetup{
    colorlinks=true,citecolor=blue, linkcolor=blue,}
\chardef\us=`\_

\begin{document}

\begin{article}
\begin{opening}

\title{Solar Flare Effects on the Earth's Lower Ionosphere}

\author[addressref={aff1,aff2},corref,email={hayesla@tcd.ie}]{\inits{L.A.}\fnm{Laura A.}~\lnm{Hayes}\orcid{0000-0002-6835-2390}}%\sep
\author[addressref={aff1,aff2},email={oharao@tcd.ie}]{\inits{O.S.D.}\fnm{Oscar S.D.}~\lnm{O'Hara}}%\sep
\author[addressref={aff2,aff1},email={sophie.murray@dias.ie}]{\inits{S.A.}\fnm{Sophie A.}~\lnm{Murray}\orcid{0000-0002-9378-5315}}
\author[addressref=aff2,email={peter.gallagher@dias.ie }]{\inits{P.T.}\fnm{Peter T.}~\lnm{Gallagher}\orcid{0000-0001-9745-0400}}%\sep
%\author{\inits{}\fnm{}~\lnm{}\orcid{}}
%\author{P.~\surname{Author-a}$^{1}$\sep
%        E.~\surname{Author-b}$^{1}$\sep
%        M.~\surname{Author-c}$^{2}$      
%       }

%   \institute{$^{1}$ First affiliation
%                     email: \url{e.mail-a} email: \url{e.mail-b}\\ 
%              $^{2}$ Second affiliation
%                     email: \url{e.mail-c} \\
%             }
\address[id=aff1]{School of Physics, Trinity College Dublin, Dublin, Ireland}
\address[id=aff2]{Astronomy \& Astrophysics Section, School of Cosmic Physics, DIAS Dunsink Observatory, Dublin D15XR2R, Ireland}

\runningauthor{L.A. Hayes et al.}
\runningtitle{Solar Flare Impacts on the Earth's Lower Ionosphere}

\begin{abstract}
Solar flares significantly impact the conditions of the Earth’s ionosphere. In particular, the sudden increase in X-ray flux during a flare penetrates down to the lowest-lying D-region and dominates ionization at these altitudes ($\approx$60\,--\,100\,km). Measurements of very low frequency (VLF: 3\,--\,30\,kHz) radio waves that reflect at D-region altitudes provide a unique remote-sensing probe to investigate the D-region response to solar flare emissions. Here, using a combination of VLF amplitude measurements at 24\,kHz together with X-ray observations from the \textit{Geostationary Operational Environment Satellite} (GOES) X-ray sensor, we present a large-scale statistical study of 334 solar flare events and their impacts on the D-region over the past solar cycle. Focusing on both GOES broadband X-ray channels, we investigate how the flare peak fluxes and position on the solar disk dictate an ionospheric response and extend this to investigate the characteristic time delay between incident X-ray flux and the D-region response. We show that the VLF amplitude linearly correlates with both the 1\,--\,8\,\AA\ and 0.5\,--\,4\,\AA\ channels, with correlation coefficients of 0.80 and 0.79, respectively. For the two X-class flares in our sample, however, there appears to be a turnover in the linear relationship, similar to previous works. 
Unlike higher altitude ionospheric regions for which the location of the flare on the solar disk affects the ionospheric response, we find that the D-region response to solar flares does not depend on the flare location. By comparing the time delays between the peak X-ray fluxes in both GOES channels and VLF amplitudes, we find that there is an important difference between the D-region response and the X-ray spectral band. We also demonstrate for several flare events that show a negative time delay, the peak VLF amplitude matches with the impulsive 25\,--\,50\,keV hard X-ray fluxes measured by the \textit{Ramaty High Energy Solar Spectroscopic Imager} (RHESSI). These results highlight the future importance of performing full spectral analysis when studying the ionospheric responses to solar flares.

\end{abstract}
\end{opening}
%-------------------------------------------------

\section{Introduction}
     \label{S-Introduction} 

% intro
Solar flares have a direct consequence on the ionized portion of the Earth's atmosphere: the ionosphere.  The increased levels of X-ray and extreme ultraviolet (EUV) emission from the Sun during a solar flare drive a rapid and significant increase of ionization over the entire dayside ionosphere \citep{tsurutani, liu2011solar, qian2011variability, yasyukevich20186}. In the lowest-lying regions of the ionosphere the electron-density profile can significantly increase over a short period of time. This in turn has important implications on radio-wave propagation over the entire radio spectrum that propagate sub- and trans-ionospherically \citep{davies1990ionospheric}. Immediate adverse space-weather effects of solar flares on the ionosphere include impacting high-frequency (HF) radio communications, GPS/GNSS navigation \citep[see][]{berdermann2018ionospheric}, air traffic control facilities, and low Earth satellite orbits. A recent adverse event occurred in September 2017 when two of the largest solar flares of the past solar cycle coincidentally took place at the same time as the Caribbean hurricanes, Jose and Irma. The large X-class flares significantly ionized the lower ionosphere resulting in radio blackouts for several hours (as the X-ray fluxes were elevated well into the flare decay phase) impacting emergency responders who were relying on HF communications at the time \citep{redmon}. Hence, building a better understanding of the solar-terrestrial relationship, particularly during solar flares, is a major focus of space-weather research. 

% solar flare effects on ionosphere - D-region most impacted
The impact on the ionosphere in response to incident flare X-ray and EUV emission is known as a sudden ionospheric disturbance (SID) and affects the various levels of the ionosphere depending on the incident wavelength \citep{mitra}. The  ionization  enhancement  can  also  perturb  the  geomagnetic field, where variations are measured in ground-based magnetometers located in the sunlit hemisphere at the time of a flare \citep[see][for a review]{curto2020geomagnetic}. These are often called a geomagnetic “crochet” or solar flare effect (sfe) and were first identified in response to the Carrington event \citep{carrington, hodgson} where a geomagnetic variation was observed in self-recording magnetometers at the time \citep{stewart}. While all regions of the ionosphere are impacted by increased solar flare ionization, the effect on the lowest-lying D-region ($\approx$60\,--\,100\,km in altitude) is most apparent. The D-region is formed and maintained during the day by solar Lyman-$\alpha$ emission (1216\,\AA) that ionizes nitric oxide (NO). However, when a flare occurs, X-rays with wavelengths $<$ 10\,\AA\ penetrate down to D-region altitudes and dominate ionization acting on the major neutral constituents including nitrogen (N$_2$) and oxygen (O$_2$) \citep{hargreaves1992solar}. This leads to a substantial increase of the electron-density profile in the D-region which in turn changes the propagation conditions of radio waves that travel through it \citep{mitra, davies1990ionospheric}. This has a particularly deleterious impact on HF radio waves that propagate through the D-region to reflect at higher altitude regions of the ionosphere. 

% VLF measurements
The response of the D-region electron-density to ionizing disturbances can be reliably probed through the observation of very low frequency (VLF: 3\--\,30\,kHz) radio signals that propagate in the waveguide formed between the Earth’s surface and the lower ionosphere. The D-region acts as a sharp reflective boundary to VLF radio frequencies, which can propagate over very large distances around the globe. In undisturbed conditions, the propagation is stable in both amplitude and phase and is characterized by a diurnal variation \citep{mcrae2000vlf, thomson2011daytime}. When a solar flare occurs, the increased ionization of the D-region changes the conductivity profile of the upper waveguide boundary for the VLF propagation, resulting in amplitude and phase variations of a measured VLF signal. Hence, measurements of narrowband VLF signals that are generated by large navigational transmitters around the world can be used to detect SIDs in the D-region ionosphere over sunlit paths.  Given that the altitude of the D-region is challenging to directly probe as it is too low for satellites and too high for stratospheric balloons and that the typical ionospheric measurement techniques, such as ionosonde and incoherent scatter, do not work well given the comparatively low electron-density at the D-region compared to the higher regions of the ionosphere (i.e. E- and F-regions), VLF remote sensing of the D-region provides a unique tool to probe this complex region and gain an understanding of the dynamical response to ionizing disturbances. 

VLF propagation studies have been utilized extensively in the past to assess the impacts of solar flares on the Earth’s lower ionosphere \citep[e.g.][]{mitra, pant, thomson05, selvakumaran}, and have similarly been used to investigate other ionospheric disturbances in the D-region such as solar proton events \citep{clilvard_2009}, $\gamma$-ray bursts \citep{inan1999}, lightning strikes \citep{inan2010survey}, solar eclipses \citep{cohen2018lower}, and even earthquakes \citep{parrot1989vlf}. To quantify solar flare impacts on the D-region, past studies have focused on the comparisons between solar X-ray flux observed in the 1\,--\,8\,\AA\ channel of the \textit{Geostationary Operational Environment Satellite} (GOES) X-ray Sensor (XRS), and measurements of VLF amplitude and phase variations \citep[e.g.][]{thomson_clivard_2001, kumar2018solar, singh2014solar,grubor2005influence}. For example, \cite{mcrae04} demonstrated that the VLF phase variation in response to a flare is proportional to the logarithm of the GOES X-ray flux, and building on this it was demonstrated that the Earth’s lower ionosphere can be used as a giant X-ray detector -- overcoming limitations of detector saturation that can happen for extremely large events \citep{thomson04}. More recently, VLF observations of solar flares have been used to develop nowcasting capabilities for X-class solar flares \citep{george}. 

By combining VLF measurements with theoretical models, the electron-density profile in the D-region can be estimated in response to solar flares \citep{wait1964characteristics, ferguson1998computer}. It has been established from a number of studies that the electron-density profile can vary by several orders of magnitude over a flaring event \citep[e.g.][]{grubor2005influence, kumar2018solar}, and can even be driven quasi-periodically in response to flaring quasi-periodic pulsations \citep{hayes2017}. Other works have focused on the time delay between the peak of incident X-ray flux and the peak of the VLF amplitude response. This time delay represents a characteristic feature of the D-region, noted as the ``sluggishness'' \citep{appleton1953note} or ``relaxation time'' \citep{mitra} of the lower ionosphere, and it is an important consequence of the electron production and loss mechanisms occurring in the lower ionosphere given the complicated D-region chemistry \citep{basak2013effective}. Together with estimates of the electron-density profile, the measured time delay can be used estimate the effective recombination coefficient \citep{basak2013effective,zigman,hayes2017}.

In this article, we present a large-scale statistical study of VLF amplitude measurements of the ionospheric D-region in response to solar flares over the past solar cycle. Our measurements probe the region of the ionosphere over the Atlantic Ocean, and our study consists of over 300 flares of X-, M-, and C- GOES class making it one of the largest statistical studies of its kind to date. In the majority of previous statistical studies, the correlation between VLF measurements and the ionizing X-ray flux were made solely with the 1\,--\,8\,\AA\ channel of GOES/XRS, and they have primarily focused on larger flares. Here, however, we have utilized both the GOES 1\,--\,8\,\AA\ and 0.5\,--\,4\,\AA\ channels and investigate how the different spectral bands affect the ionospheric VLF response. In particular, we focus on the relationship between the peak X-ray fluxes and the VLF amplitude responses and perform a detailed study of the time delays between the different X-ray spectral bands. Furthermore, we investigate the relationship between the location of the flares on the solar disk and the D-region response, which has not yet been studied for these ionospheric altitudes. In Section~\ref{vlf_desc} of this article the VLF monitoring setup used in this study is described, and we detail our flare selection criteria and the properties used in our statistical study. The results of the statistical study are presented in Section~\ref{stat_study}, and a discussion and conclusion are provided in Section~\ref{conclusion}.

\section{Observations and Data Analysis}
\label{vlf_desc}

\begin{figure}[h!]
  \centering
  \includegraphics[width =0.9\textwidth]{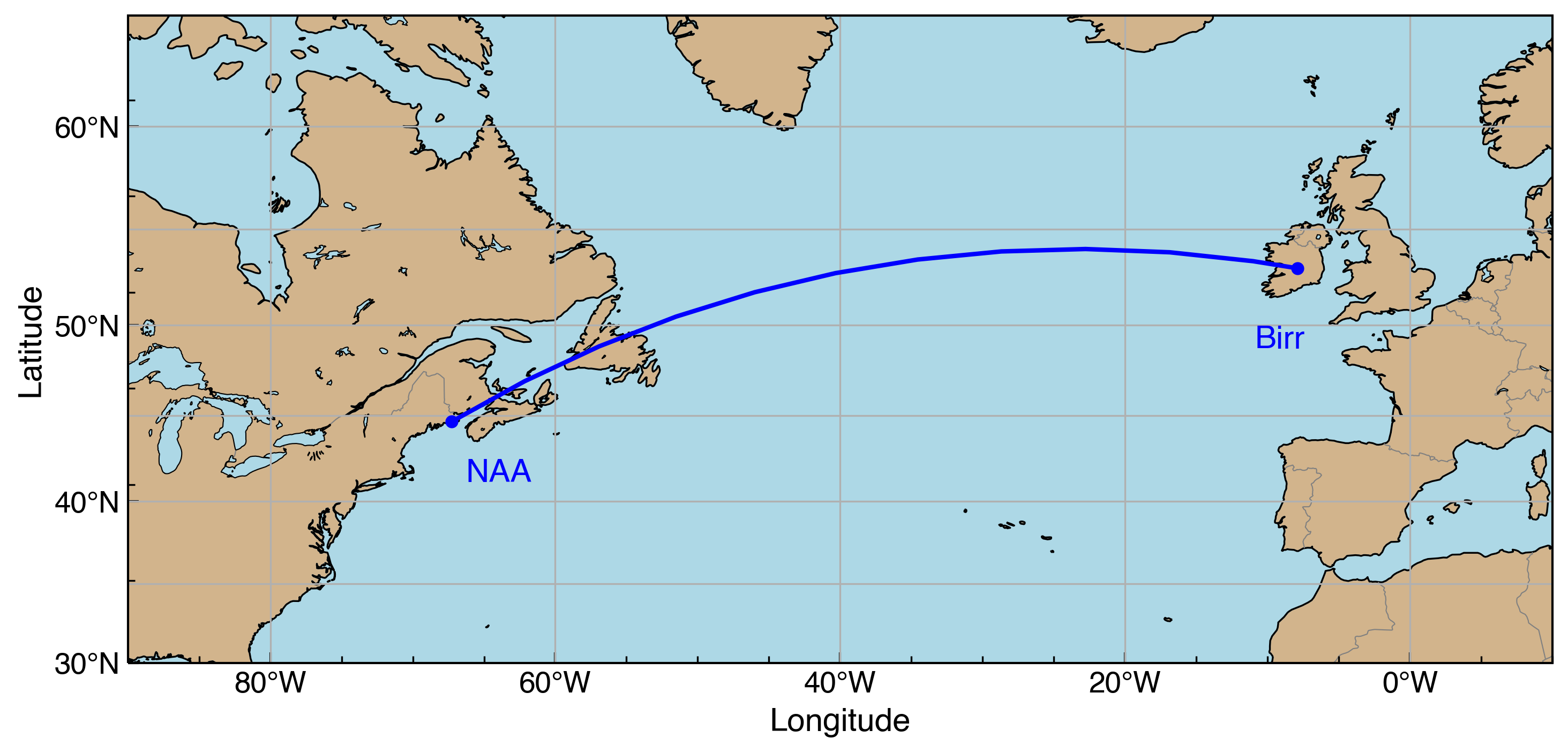}
  \caption{The VLF propagation path from the VLF transmitter NAA in Cutler, Maine, USA, to the VLF receiver located at the Rosse Observatory in Birr, Offaly, Ireland. The line connecting the location points represents the great circle path. Measurements at Birr provide ionospheric conditions in the lower D-region over this path.}
  \label{map_path}
\end{figure}

The VLF recording system used in this study consists of a receiving antenna and a \textit{Stanford Sudden Ionospheric Disturbance Monitor} \citep{scherrer} that is located at the Rosse-Solar Terrestrial Observatory at Birr Castle Demesne, Co. Offaly, Ireland (53.1$^{\circ}$N, 7.9$^{\circ}$W) \citep{zucca2012observations}. The receiving antenna is a magnetic wire-loop antenna that has a diamond shape frame (1.42 m side length) that holds a series of wire loops (23 turns). The signal to the antenna is fed into the SID monitor, which has the ability to record narrow-band VLF signal amplitude at a particular frequency. Our SID monitor is tuned to the US naval communications transmitter (call-sign NAA) in Maine, US  (44.6 $^{\circ}$N, 67.2$^{\circ}$W), that operates at 24\,kHz. This VLF propagation path has a great circle distance of $\approx$5,320~km largely over the Atlantic Ocean, and it is illustrated in Figure~\ref{map_path}. Observations from  this  VLF  receiver  have been  used  in  past  studies  of  solar  flares  effects  on the Earth’s ionosphere \citep{hayes2017}.

\begin{figure}[h!]
  \centering
  \includegraphics[width =0.9\textwidth]{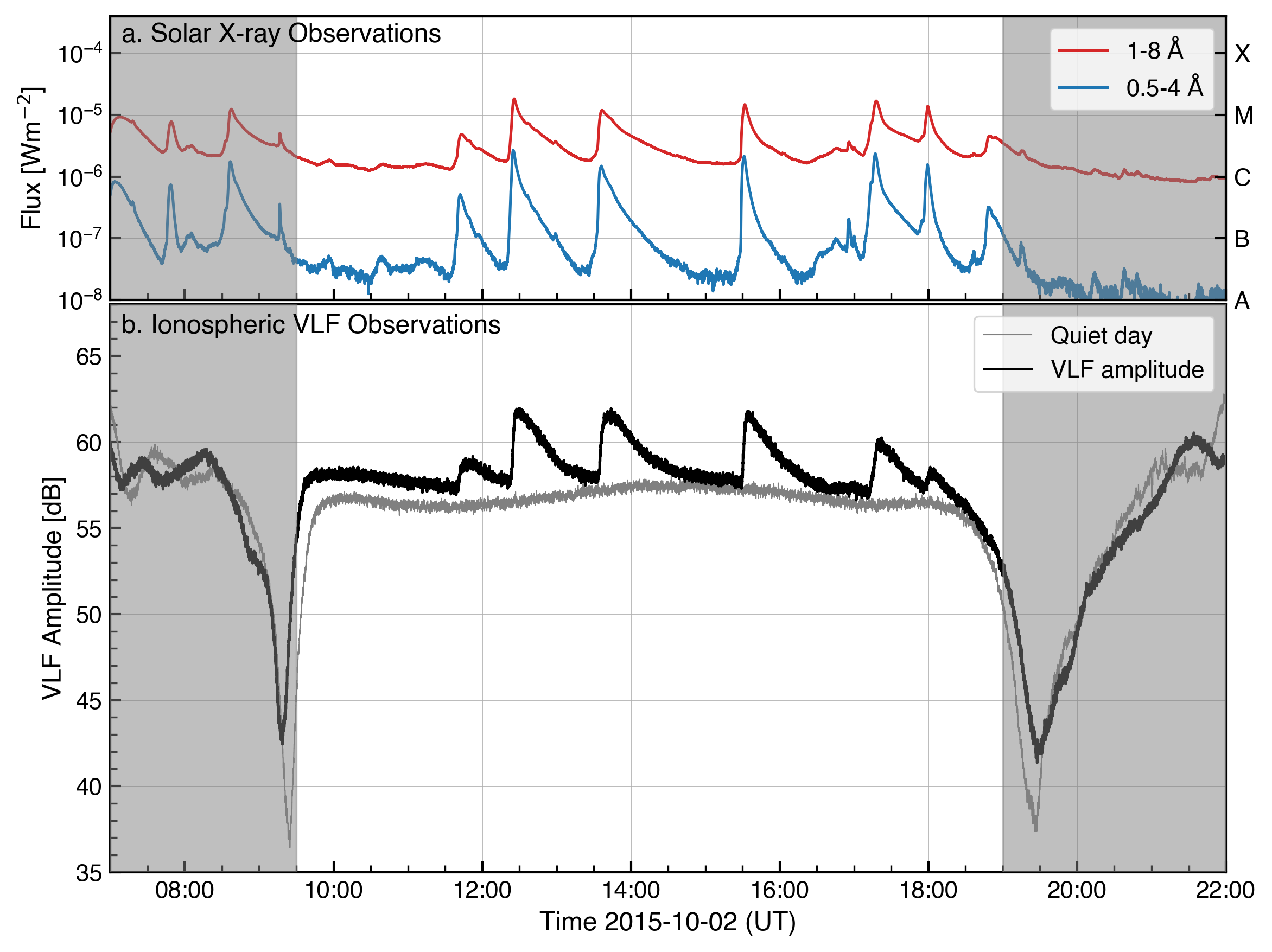}
  \caption{An example of how VLF measurements can be used to detect solar flares in the ionospheric response. Panel \textbf{a} shows the GOES/XRS measurements of solar X-ray flux in 1\,--\,8\,\AA\ and 0.5\,--\,4\,\AA. Panel \textbf{b} shows the VLF amplitude measurements from 2 Oct 2015 (black), and from a solar quiet day (i.e. no flares) from several days previous to compare (gray). The shaded regions demarcate the nighttime.}
  \label{example_day}
\end{figure}

An example of a day of observations for the VLF signal amplitude during an active flaring day on 2 October 2015 is shown in Figure~\ref{example_day}. The top panel shows solar X-ray flux from the GOES/XRS observations in the two broadband soft X-ray wavelengths, 1\,--\,8\,\AA\ (red) and 0.5\,--\,4\,\AA\ (blue). Here, several flares can be identified. The bottom panel shows the amplitude of the received VLF signal. Clearly, during the daytime, solar flares can readily be detected in the variations of the VLF amplitude (black curve), reflecting the enhancements in the D-region electron-density caused by the increased flare X-ray fluxes. The gray shaded region demarcates night time. The signal sharply falls and rises both during sunrise and sunset, and it follows a diurnal pattern during the day \citep{mcrae2000vlf}. For comparison the gray curve shows the VLF signal amplitude during a quiet day in which no solar flare occurs. The increase in VLF amplitude in response to the flare ionization is due to the fact that the increased electron-density profile in the D-region provides a sharper upper boundary for the propagating VLF wave, hence reducing attenuation \citep{thomson_clivard_2001}.

The availability of the VLF measurements used in this study span from 2012 to late 2018 with some large data gaps corresponding to times when the receiver was offline, or the antenna was disrupted, including for example, the majority of 2014.  For the purpose of our analysis we focused on solar flare events greater than a C1.0 GOES class ($>1\times10^{-6}$~Wm$^{-2}$) that occurred during sunlit hours over the VLF propagation path from the transmitter NAA to the SID receiver in Birr (i.e. Figure~\ref{map_path}). We used the National Oceanic and Atmospheric Administration/Space Weather Prediction Center (NOAA/SWPC) GOES flare list to identify events over the time period for which we have reliable VLF data between 2012 and 2018. The GOES flare list provides the start, peak, and end times of each X-ray flare event, as well as the peak flux. It however does not always include the location of the flare on the solar disk. As we were interested in studying the relationship between the location of the flare on disk to its geophysical impact on the ionosphere, we accessed the flare locations from the SSW Latest Events list (see \url{www.lmsal.com/solarsoft/latest_events_archive.html}), which determines flare positions based on imaging data from the \textit{Atmospheric Imaging Assembly} (AIA). We then cross-referenced this flare list with the availability of our VLF data and only considered flare events in which there was a noticeable VLF amplitude response at the peak time of the flare. We then further removed events that occurred close to the sunrise--sunset diurnal changes. Following these selection criteria, a total of 334 flare events were left on which to perform our statistical study: 261 C-class flares, 71 M-class flares, and 2 X-class flares.

For each flare, we used the start and end times from the GOES flare list to define the flare time range to investigate the ionospheric response with the VLF amplitude measurements (see \url{www.swpc.noaa.gov/products/goes-x-ray-flux} for definition of start and end times). We then subtracted a pre-flare background from the VLF measurements to quantify the increased VLF amplitude response [$\Delta A$] in response to the ionizing flare X-ray flux. This procedure is needed because D-region ionization experiences both diurnal and seasonal effects that vary the background VLF amplitude. Hence we only want to identify the amplitude variations caused by the flare ionization to get the true induced response. The background was determined as the VLF amplitude at the start time of flare as defined by the GOES catalog, such that $\Delta A(t) = A(t) - A(t_{\mathrm{goes\_start}})$. Similarly, we subtracted the pre-flare levels in both channels of the GOES/XRS measurements for each flare. This is similarly required as the background-flux levels in the GOES channels can be quite high when the Sun is active. To perform a statistical analysis we determined for each flare the peak flare flux in both GOES channels, the VLF amplitude increase [$\Delta A$] induced by the flare, and the peak times of both the GOES fluxes and the VLF amplitude. In some cases, the VLF data had quite a low signal-to-noise ratio, and hence to determine the maximum value and timing we smooth the data over the time range of the flare using a Savitzky–Golay filter \citep{savitzky1964smoothing} with a two-minute window.

\begin{figure}[h!]
  \centering
  \includegraphics[width =0.8\textwidth]{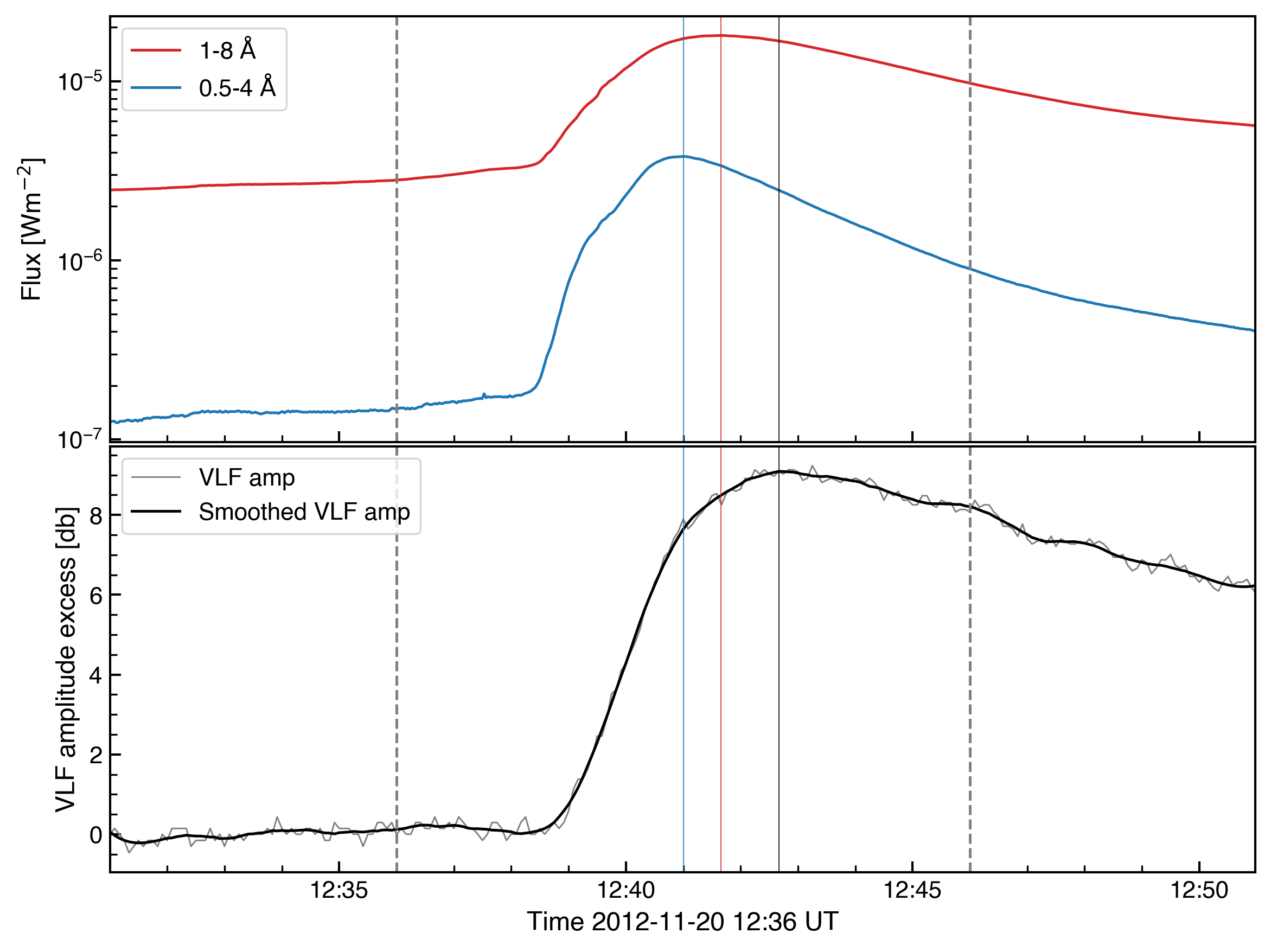}
  \caption{An example of the analysis performed on each flare that we used in this statistical study. The top panel shows both the GOES channels (1\,--\,8\,\AA\ and 0.5\,--\,4\,\AA). The vertical dashed lines show the GOES defined start and end times of the flare, and the red and blue vertical lines show the peak of the 1\,--\,8\,\AA\ and the 0.5\,--\,4\,\AA\ channels, respectively. The bottom panel shows the background removed VLF amplitude in gray, and the smoothed signal in black. The black vertical line shows the VLF amplitude peak.}
  \label{example_ana}
\end{figure}

An example of one flare from our sample is shown in Figure~\ref{example_ana}. The top panel shows the GOES 1\,--\,8\,\AA\ and 0.5\,--\,4\,\AA\ channels in red and blue, respectively, and the bottom panel shows the corresponding background-subtracted VLF amplitude in gray, and the smoothed amplitude in black. The vertical gray dashed lines mark the defined GOES start and end times of the flare, the red and blue vertical dashed lines show the peak of the flare in the 1\,--\,8\,\AA\ and 0.5\,--\,4\,\AA\ channels, respectively, and the black vertical line marks the peak of the VLF amplitude. The characteristic time delay between the peak of the X-rays and the VLF amplitude is clearly demonstrated here. For each flare, these values are calculated and used to determine the associated VLF amplitude response for each flare, and the respective time delays between the X-ray and VLF amplitude peaks. These are then utilized in a statistical analysis to understand the relationship between the incident flare fluxes and the D-region response.

\section{Statistical Study Results} %%%%%%%%%%%%%%%%%%%%%%%%%%%%%%%%%%%%%%%%
      \label{stat_study}      

\subsection{VLF Amplitude Response to Flare X-Ray Fluxes}

\begin{figure}[h!]
  \centering
  \includegraphics[width=\linewidth]{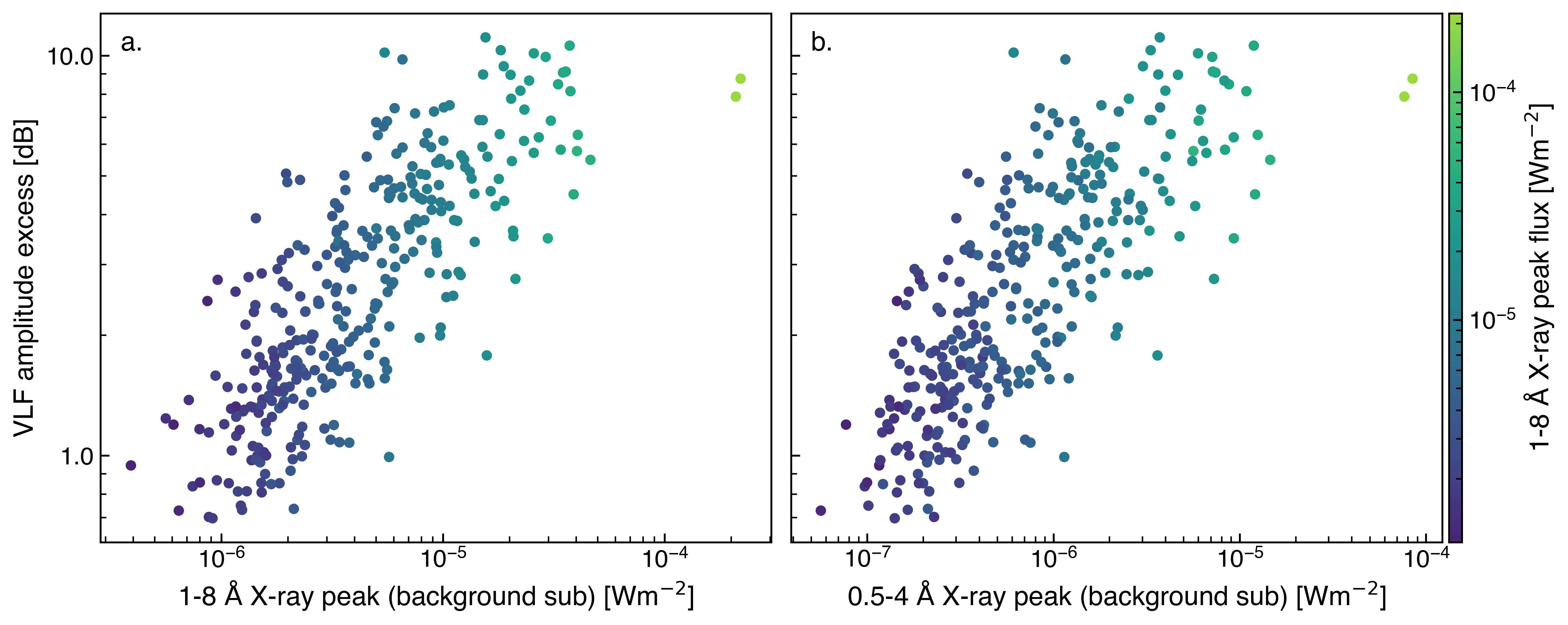}
  \caption{Correlation of the excess VLF amplitude and GOES X-ray flux for both the 1\,--\,8\,\AA\ channel \textbf{(a)} and the 0.5\,--\,4\,\AA\ channel \textbf{(b)}. The X-ray fluxes have had their pre-flare flux levels subtracted. The color-scale represents their peak X-ray flux in the 1\,--\,8\,\AA\ (without background subtraction), hence illustrating their GOES class.}
  \label{peak_flux}
\end{figure}

For each of the flares in our sample, we first analyze the relationship between the peak X-ray fluxes (background-subtracted) in the 1\,--\,8\,\AA\ and 0.5\,--\,4\,\AA\ channels and the excess VLF amplitude variations. The results of this is shown in Figure~\ref{peak_flux}~a and b for 1\,--\,8\,\AA\ and 0.5\,--\,4\,\AA, respectively. The color of the points represent their peak flux in the 1\,--\,8\,\AA\ (not background-subtracted) to reflect their respective GOES classes. A strong correlation is clearly identified in both, with higher flare fluxes resulting in a larger VLF amplitude response. Calculating the Spearman rank correlation coefficient of the relationships, we find that it is 0.80 and 0.79 for 1\,--\,8\,\AA\ and 0.5\,--\,4\,\AA\ channels, respectively.

Given that the VLF amplitude increases reflect larger changes in the electron-density profile in the D-region, these results demonstrate that the higher the peak of the incident X-ray flux, the larger the ionospheric disturbance. The linear relationship between the two appears to hold up to the GOES M5-class, after which there appears to be a turnover, where this linear trend changes to be more of a sigmoid shape. However, given that there are only two X-class flares in our sample, the question arises as to whether these are two outlier points that lie within the scatter of the linear relationship, or whether there is a turnover point for which the amplitude does not continue to increase linearly with increasing flux. The observation of the turnover has however been reported in earlier works \citep[e.g.][]{mcrae04} that have looked at this relationship in regards to the ionospheric Wait parameters \citep{wait1964characteristics}, [$H'$] and [$\beta$], which are parameters that characterize the D-region. $H'$ is the ionospheric ``reflection'' height and $\beta$ is the rate of increase of electron-density with height, or ``sharpness'' of the D-region. It has been demonstrated that the $\beta$-parameter reaches an upper limit at a certain level, and given that the VLF amplitude measurements are more representative of $\beta$ (as compared to the phase which relates closer to $H'$) this may help us to interpret our result here. However, as noted, we unfortunately only have two X-class flares in our sample and future observations will allow us to investigate this relationship in more detail. Another important insight from Figure~\ref{peak_flux} is that even within the linear relationship between the peak X-ray fluxes and the ionospheric response there is large scatter. Many background factors may affect this, such as prior ionospheric conditions, time of day and year, however by background-subtracting our VLF signal amplitude large trends that would impact this would have been removed. The scatter may also be indicative of different spectral components of the flares having an impact on the ionospheric response. For example, flares with similar soft X-ray components (i.e. GOES classes) can have significantly different hard X-ray spectra and hence could play a role in producing the scatter evident here.

\subsection{Relationship to Heliographic Location of Flare} \label{stat_loc}

\begin{figure}[h!]
  \centering
  \includegraphics[width =0.99\textwidth]{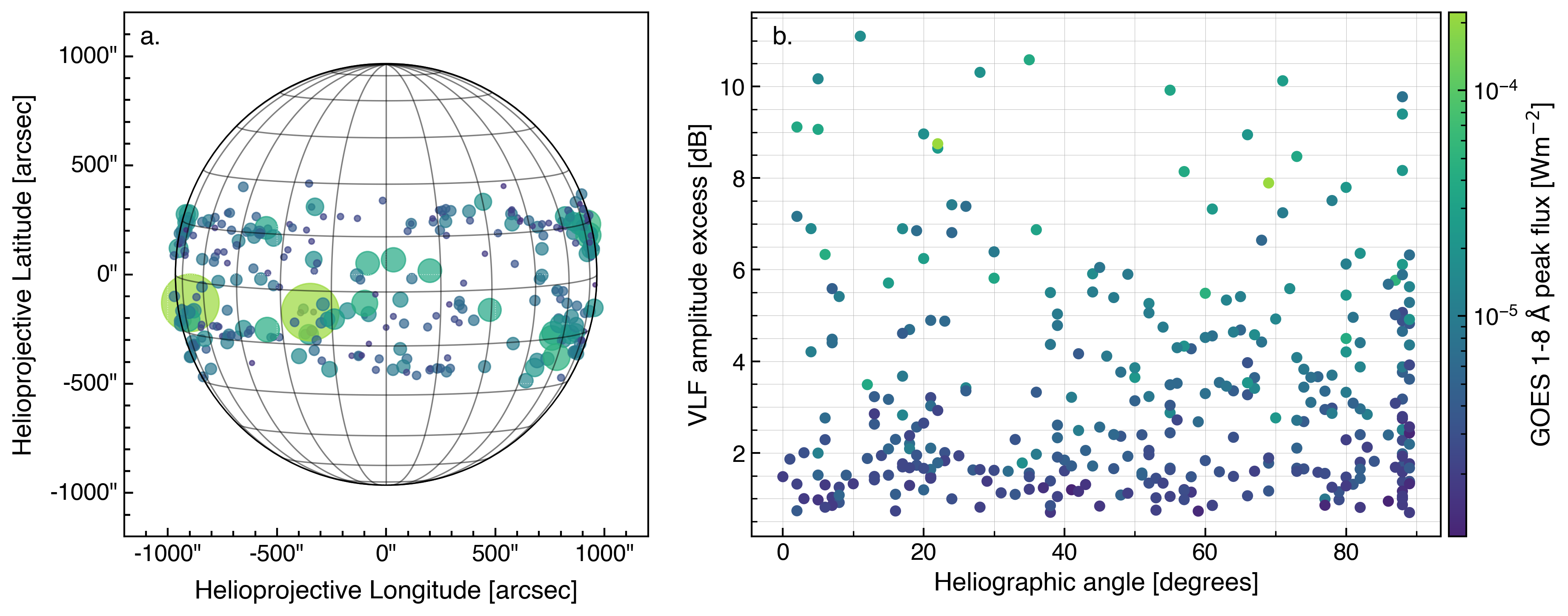}
  \caption{\textbf{a:} The location of the sample of flares from our statistical sample on disk. The color and size represent their GOES class. \textbf{b:} A scatter plot of the VLF amplitude and the heliographic position of the incident flares. Clearly no relationship is found, indicating that the position of a flare on disk does not impact the geoeffectiveness of a flare on the lower ionosphere. The appearance of many flares at around 90$^{\circ}$ is due to projection effects.}
  \label{location_flares}
\end{figure}

An important relationship to investigate is how the location of the flare on the solar disk relates to the geophysical impact on the lower ionosphere. In the upper regions of the ionosphere, such as the E- and F-regions, it has been demonstrated that the location of a solar flare on the disk of the Sun dictates how the ionospheric system will respond to the flare \citep{zhang2011impact, qian2019solar}. However this has not yet been fully investigated for the lower ionosphere. To investigate this for the D-region, we compare the locations of each flare in our sample to the excess VLF amplitude response. To illustrate the locations of all the flares, we have plotted their locations in helioprojective coordinates on the solar disk in Figure~\ref{location_flares}~a, where the sizes and color of each point represents the GOES class. Figure~\ref{location_flares}~b shows the relationship between the heliographic location of the flare to VLF amplitude response. There is clearly no relationship, which is confirmed with a Spearman rank correlation coefficient of 0.02. This demonstrates that the location of the flare on the solar disk does not have impact on 
how the lower D-region ionosphere will respond.

This difference between the relationships found at higher ionospheric altitudes and the D-region can be explained in terms of the center-to-limb variation of certain EUV wavelengths. At higher altitudes in the ionosphere, the solar EUV flux dominates the ionization, whereas in the D-region it is the solar X-ray flux that dominate during a flare. The differences in optical thickness of the EUV and X-ray fluxes is hence playing an important role here. Some EUV emissions are optically thick, particularly chromospheric lines, and hence they show a strong center-to-limb variation such that flares that occur closer to the solar limb have a reduced irradiance at Earth compared to those that occur near the disk center. This is due to the fact that those flares near the limb have a longer optical path (increasing column mass along the line-of-sight) and are more absorbed in the solar atmosphere. X-rays, on the other hand, are optically thin, and do not show this variation. This can thus explain how the D-region response to ionizing flare emission does not show the relationship of heliographic position. These results also further prove that it is the X-ray emissions that are dominant during flares over the Lyman-$\alpha$ emission, which dominates the ionization of the D-region during the day during non-flaring times. Lyman-$\alpha$ emission has a strong center-to-limb variation \citep[see Figure 6 of][]{milligan2021solar} as it is an optically thick line, and hence if the excess Lyman-$\alpha$ emission during a flare were to have a significant impact, we should also see this relationship in our Figure~\ref{location_flares}~b. This work further supports the study of \cite{raulin2013} who found that the ionospheric response in the D-region does not depend on the presence of enhanced flaring Lyman-$\alpha$ emission for a small sample of seven flares.

\subsection{Time Delay Between Peak X-ray and VLF Amplitude Response}

The electron-density in the D-region is controlled by the balance between photoionization
and electron loss processes. The time delay that is noted between the VLF amplitude and the X-ray flux is the inherent chemical reaction timescale of the D-region which results from recombination (electron loss) processes \citep{appleton1953note, mitra}. At D-region altitudes, the dominant loss processes are electron--ion, ion--ion and three--body recombination \citep{whitten1961model, mitra, mitra1972ionospheric, zigman}. These processes together can be quantified in terms of the effective recombination coefficient [$\alpha_{\mathrm{eff}}$] that embodies the combined effect \citep{whitten1965effective}. The time delay found between the VLF amplitude (and phase) peak and the incident X-ray peak is commonly found to be on the order of minute timescales \citep{zigman}. As a measurable quantity, it has been used in the past together with the electron continuity equation to determine the $\alpha_{\mathrm{eff}}$ \citep[e.g.][]{zigman, basak2013effective, hayes2017}.

\begin{figure}[h!]
  \centering
  \includegraphics[width=\linewidth]{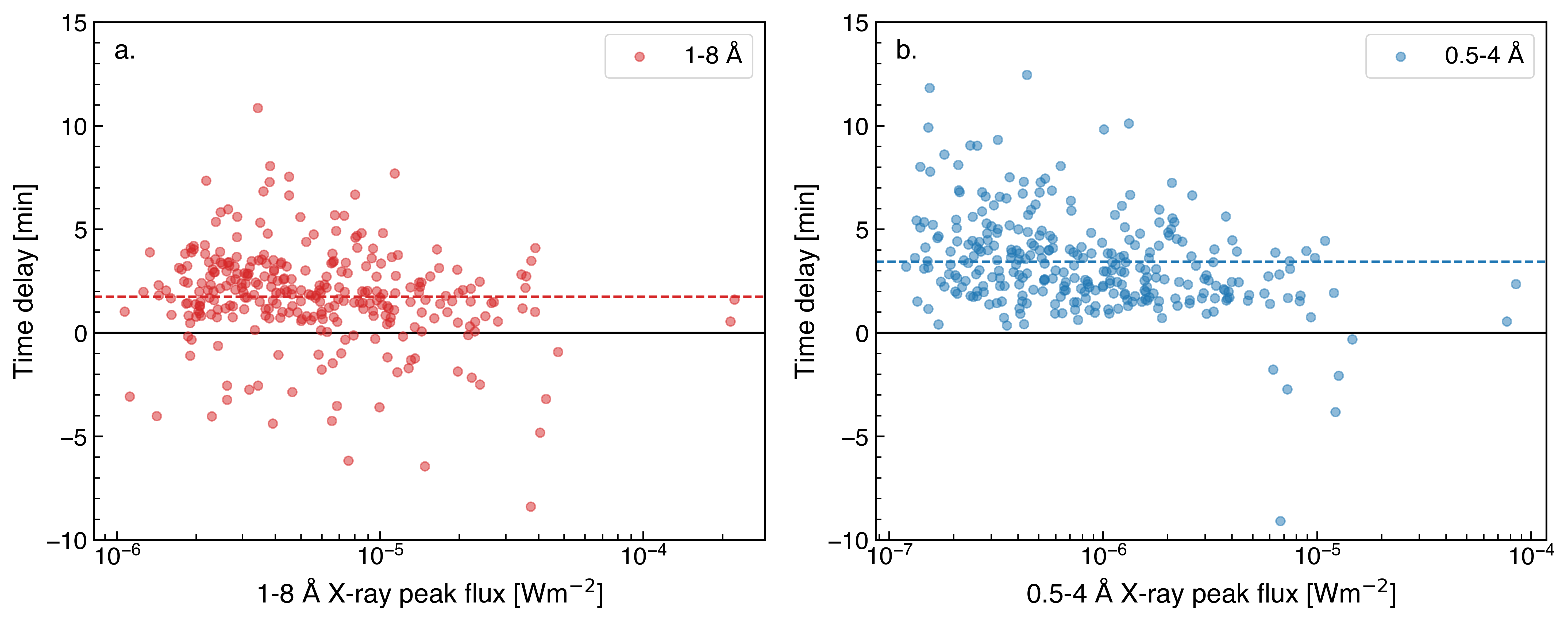}
  \caption{The relationships of the time delay between the X-ray flux the VLF amplitude and as a function of peak flux for both the GOES 1\,--\,8\,\AA\ channel \textbf{(a)} and the 0.5\,--\,4\,\AA\ channel \textbf{(b)}. The horizontal black line marks the zero value (i.e. no time delay), and the horizontal dashed lines mark the mean of the distributions.}
  \label{time_delay}
\end{figure}

To build towards a better understanding of the time delay of the ionospheric response in a statistical manner, we analyze these measurements from our sample of solar flares. Previous studies have typically focused on the time delay estimated between the VLF signal peak and only the 1\,--\,8\,\AA\ channel, however it is important to compare the time delay in terms of different X-ray spectral bands. Here, we used both GOES/XRS channels and investigate the time delays between X-ray fluxes and the VLF amplitude response to assess how the incident wavelengths play a role. For each flare event, the time delay was determined between the peak of the 1\,--\,8\,\AA\ and 0.5\,--\,4\,\AA\ channels and the relative VLF amplitude peak (see Figure~\ref{example_ana} for example). These time delays are plotted in Figure~\ref{time_delay} for the 1\,--\,8\,\AA\ channel in panel a, and the 0.5\,--\,4\,\AA\ channel in panel b as a function of their respective peak X-ray flux values. There does not appear to be a clear trend between X-ray peak flux and time delay, and the points are quite scattered. This scatter is most likely due to factors other than the X-ray wavebands here playing a role in the response. While the background subtraction should remove most  of the large scale effects, such as the time of day, time of year, and pre-flare ionospheric conditions, their influence may still contribute to the scatter observed. Other factors, in particular the spectral components of the flares are most likely a cause. A small negative correlation is found of -0.20 for the 1\,--\,8\,\AA\ channel and -0.31 for the 0.5\,--\,4\,\AA\ channel such that for higher X-ray fluxes there is a smaller time delay. The black horizontal line is at zero time delay. For negative values (i.e. points below this black line), this means that the VLF amplitude peaked \textit{before} the X-ray flux. It is important to note that for several events the VLF amplitude peaks before the GOES 1\,--\,8\,\AA\ channel but not before 0.5\,--\,4\,\AA\ channel. This has important implications suggesting that it is the higher energy (``harder'') X-rays that are driving the ionization.

\begin{figure}[h!]
  \centering
  \includegraphics[width=0.5\linewidth]{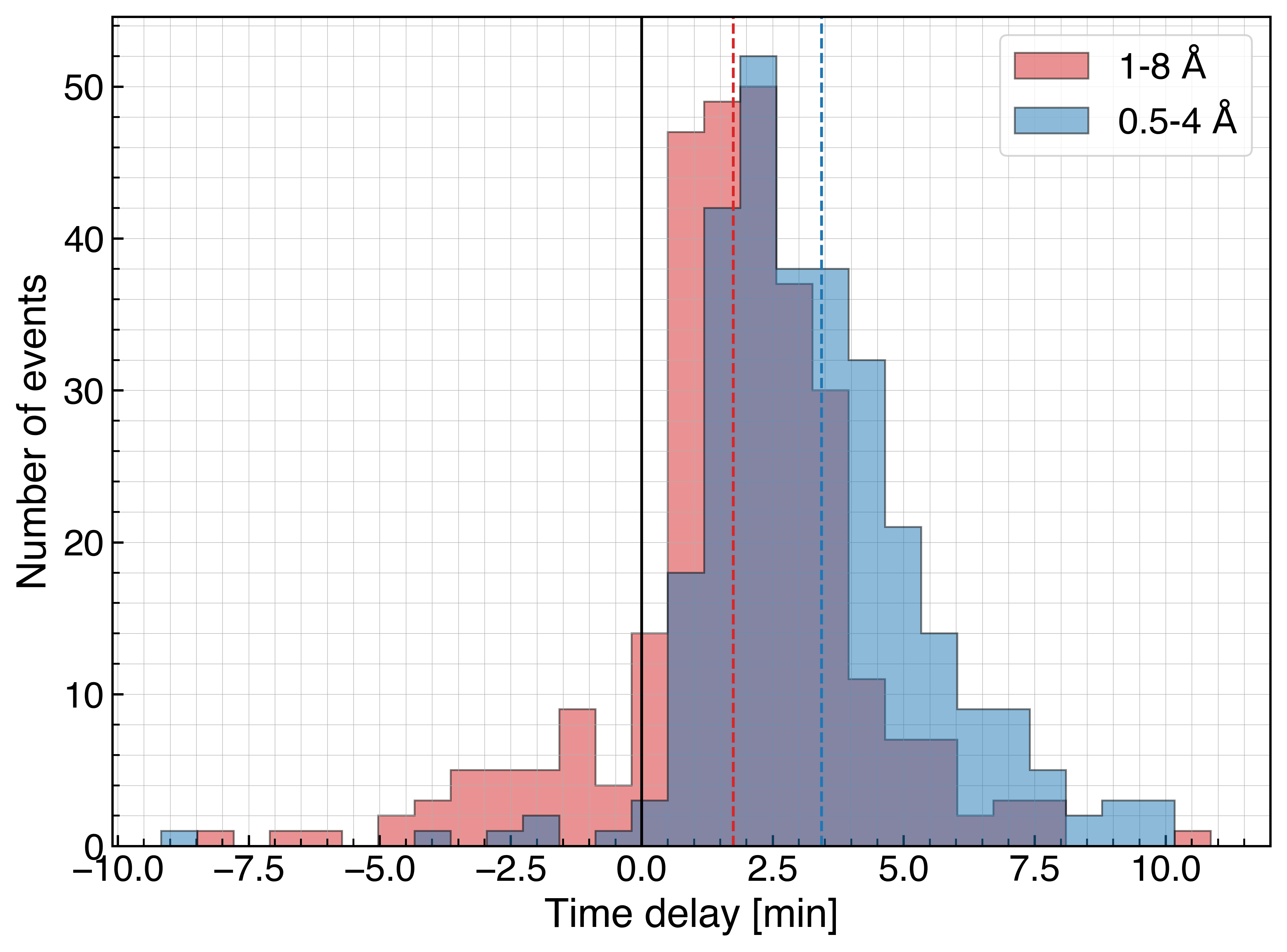}
  \caption{The distribution of the time delays between the VLF peak and the X-ray 1\,--\,8~\AA\ (red) and 0.5\,--\,4~\AA\ (blue) for all events in our sample. The vertical dashed lines mark the mean values of both distributions.}
  \label{time_delay_dist}
\end{figure}

The distributions of the measured time delays between the X-ray fluxes and the VLF amplitude response are shown in Figure~\ref{time_delay_dist}. This again clearly demonstrates the differences between the two X-ray channels. The mean values of the distributions are marked by the vertical dashed lines in their respective colors, and are 1.7\,minutes for the 1\,--\,8\,\AA\ channel and 3.4\,minutes for the 0.5\,--\,4\,\AA\ channel. This figure highlights again that the time delay between different X-ray spectral bands can be quite different. Here we only focused on the two broadband GOES X-ray channels, but ideally a comparison should be made between the full X-ray spectra over a large energy range to determine the dominant spectral components that ionizes the D-region. 

\begin{figure}[h!]
  \centering
  \includegraphics[width=0.8\linewidth]{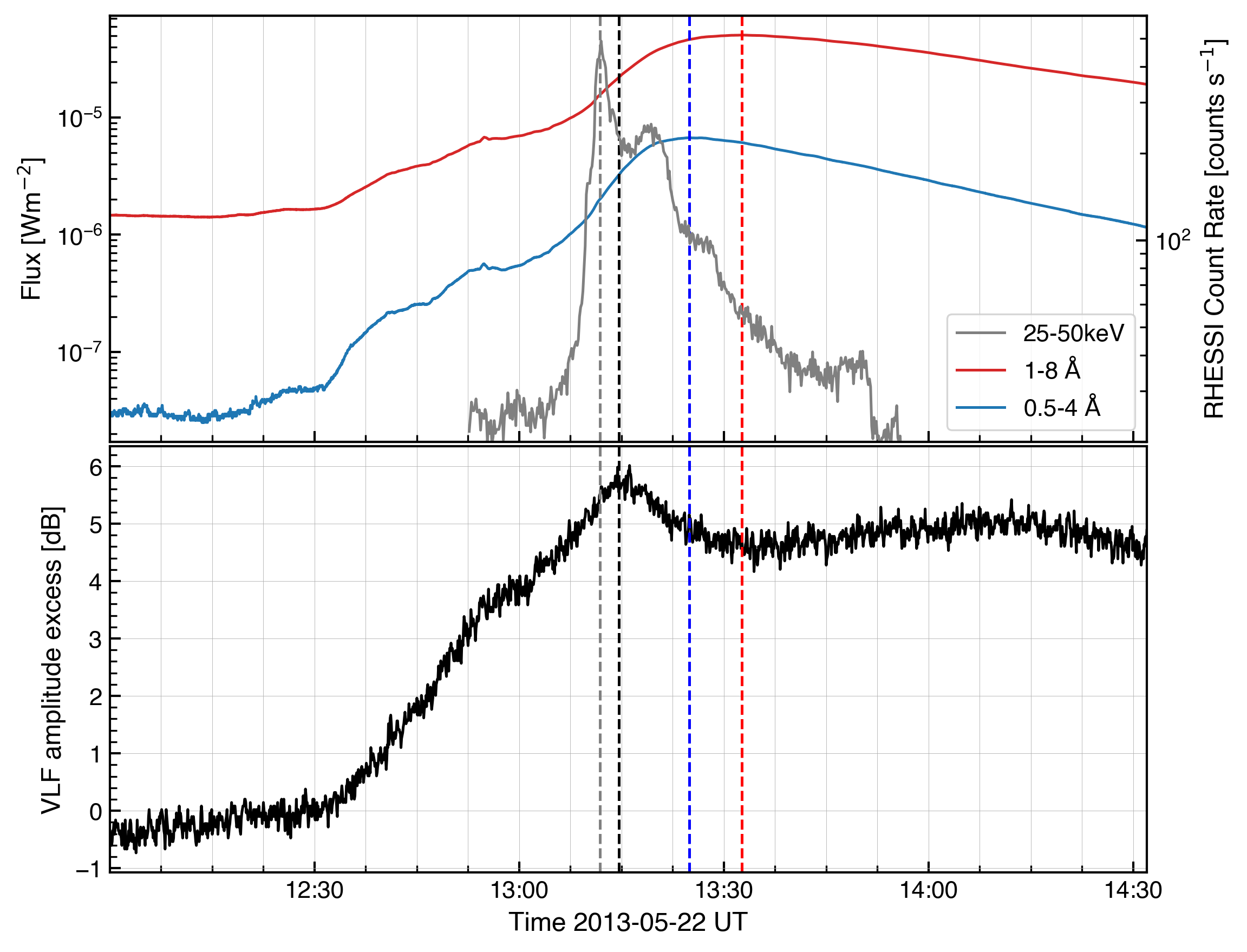}
  \caption{An M-class solar flare from our sample in which the VLF amplitude peaked significantly before the peaks in either GOES channels. The top panel shows both GOES channels in red (1\,--\,8\,\AA) and blue (0.5\,--\,4\,\AA), and their respective peak times marked by the vertical dashed lines. The gray curve plots the RHESSI 25\,--\,50\,keV data, clearly demonstrating hard X-ray signatures associated with non-thermal emission. The bottom plot shows the VLF amplitude response, where it can clearly be seen to peak before the GOES X-rays. It interestingly matches the higher energy hard X-ray RHESSI time-profile.}
  \label{fig:time_delay_example}
\end{figure}

From both Figure~\ref{time_delay} and \ref{time_delay_dist} there are clearly several events in our sample for which the VLF amplitude peaks before either of the GOES/XRS channels. This raises an important question as to what is causing the induced D-region response in these cases, and it is suggestive that perhaps harder non-thermal X-rays are playing a role. To further investigate this, we chose the M5.0-class solar flare from 22 May 2013 that showed the largest time gap between the VLF response peak and the GOES/XRS fluxes to study in more detail with higher energy X-ray observations from \textit{Reuven Ramaty High Energy Solar Spectroscopic Imager} \citep[RHESSI:][]{lin2003reuven}. The results of this are shown in Figure~\ref{fig:time_delay_example}. The top panel shows the X-ray fluxes in both GOES/XRS channels and the RHESSI 25\,--\,50\,keV observations. As expected, the hard X-ray 25\,--\,50\,keV observations peak before the soft X-ray GOES/XRS observations, reflecting the non-thermal emission associated with accelerated electrons during the flare impulsive phase. The bottom panel shows the VLF amplitude excess in response to the flare. Clearly, the peak in the VLF amplitude occurs before the GOES/XRS measurements. However, the large VLF peak matches closely with the impulsive hard X-ray RHESSI observations with a time delay as expected between the peak of the X-rays and the peak of the ionospheric response. In the case of this flare, it appears that the hard X-ray emission is dominating the ionization of the D-region during the impulsive phase of the flare. Following the impulsive phase, the VLF amplitude shape follows to match the time profile of the soft X-rays. This example further demonstrates a need to investigate how the spectral components of a solar flare impact the ionospheric response. For the remaining flares that peak in the VLF amplitude before both GOES/XRS channels, we similarly find that the hard X-ray observations match the early impulsive peak in the VLF amplitude. For events for which hard X-ray observations were not available, we used the derivative of the GOES 1\,--\,8\,\AA\ channel as a proxy for the hard X-ray emission following the Neupert effect \citep{neupert}.

\section{Discussion and Conclusion} %%%%%%%%%%%%%%%%%%%%%%%%%%%%%%%%%%%%%%%%
      \label{conclusion} 
      
We present a statistical analysis of 334 flares and their impacts on the ionospheric D-region over the past solar cycle. By using a combination of observations with the two broadband X-ray channels, 1\,--\,8\,\AA\ and 0.5\,--\,4\,\AA\ of GOES/XRS and VLF amplitude measurements of the D-region over the Atlantic ocean we have been able to study the relationship between the solar flare characteristics and their impact on the lower ionosphere. Focusing on the comparison between X-ray fluxes and VLF amplitude increases, we found that the response of the D-region increases with increasing X-ray flux in both XRS channels. The VLF amplitude enhancement induced by solar flares can be explained in terms of the electron-density profile enhancement in the D-region caused by the significantly increased ionization by the incident flare X-ray flux. Our results are similar in form to those presented in past statistical studies, and the turnover shape for larger X-ray flares is in comparison with those such as \cite{mcrae04} and \cite{kumar2018solar}.

Our statistical analysis of the time delay between the X-rays fluxes and the VLF amplitude have important implications for understanding how the D-region responds to different X-ray spectral components. Our results support that of \cite{nina2018analysis} for which they illustrated in a detailed analysis of single flare that the calculations of the time delay between the electron-density maximum and the incident-radiation maximum depend on the considered flux and that choosing one energy range arbitrarily for a given flare does not provide proper information. Our results also imply that the GOES 1\,--\,8\,\AA\ channel alone should not be used when considering time delays between the ionospheric response, or used solely in the calculation to determine the effective recombination coefficient using calculated time delays. This is also particularly important when modeling the lower ionosphere in response to solar flares \citep{basak2013effective, chakraborty2020numerical}, and our results suggest that the full X-ray spectra should be considered when performing such analysis \citep[e.g.][]{palit2015theoretical}.

We find that there is no dependence of the ionospheric response to the position of a solar flare on the solar disk, meaning that flares near the solar limb have similar effects on the D-region to those that occur at the disk center. This is in contrast to the higher altitude ionosphere (F-region) which exhibits a stronger response to flares that occur at disk center given that the response is driven by optically thick EUV emissions. The fact that there is no dependence of the lower ionospheric response to the position of the flare on the solar disk further strengthens the results of \cite{raulin2013} for which they were unable to detect a D-region response to flare Lyman-$\alpha$ emission. Our results suggest that X-ray emission plays the dominant role in ionization of the D-region during a flare and that Lyman-$\alpha$ has a negligible effect on the D-region during this time. It should be acknowledged however that this does not necessarily mean that Lyman-$\alpha$ emission does have an impact on the Earth’s ionosphere during a solar flare. For example, an important recent study by \cite{milligan2020lyman} showed that for a large impulsive flare, the Lyman-$\alpha$ emission closely correlated with a crochet signature in ground-based magnetometer data suggestive of flare-induced currents in the E-region. Their results demonstrated that Lyman-$\alpha$ may have an important impact on the ionosphere at higher altitudes and highlights the need to further investigate how the contribution of different spectral components of a flare influence the different ionospheric regions. Flares with the same GOES magnitude (i.e. same soft X-ray peaks) can have significantly different X-ray spectral components and EUV emission spectra. This in turn will have a different impact on the Earths lower ionosphere. The results of this article have highlighted that to further understand the geophysical impact of solar flares on the lower ionosphere, further detailed studies of the spectral components of flares and their relationship to the different ionospheric regions is needed.

As illustrated in this article, remote sensing using VLF radio-wave propagation measurements provides a cost-effective technique to probe the physical processes in the D-region ionosphere in response to solar flare radiation. In this work, we focused on a single frequency, narrowband observation probing the region of the ionosphere over the Atlantic. In preparation for Solar Cycle 25, we have now installed two new VLF receiving systems in Ireland with upgraded Stanford SuperSID monitors that have the ability to receive a spectrum of frequencies allowing several VLF stations to be monitored at once. The measurements from a number of different transmitters from both an eastern and western direction will provide longer daytime coverage for which flares can be detected, and furthermore will allow for us to probe different geographical regions of the ionosphere. The sites of this new VLF receiving systems are at Birr, Ireland (the same location as the observatory used in this study), and at Dunsink Observatory in Dublin, Ireland. At both of these locations are co-located ground-based magnetometers, which will provide the needed measurements to study the geomagnetic effects on the ionosphere in response to solar flares and compare them with VLF amplitude measurements. Moreover, with the advent of the new X-ray data now currently taking measurements in the 4\,--\,150\,keV range available from the \textit{Spectrometer Telescope for Imaging X-rays} (STIX) on Solar Orbiter \citep{krucker2020spectrometer}, detailed X-ray spectrum measurements will also be available for future analysis of solar flare events. As the solar activity continues to increase over the next several years, our increased observational coverage will allow us to build upon the results of this article and perform a more detailed analysis of the geophysical impacts on the lower ionosphere in response to solar flare emission.

%%%%%%%%%%%%%%%%%%%%%%%%%%%%%%%%%%%%%%%%%%%%%%%%%%%%%%%%%%%%%%%%%%%%%%%%%%%
\begin{acks}
 L.A. Hayes and O.S.D. O'Hara. are supported by Air Force Office of Scientific Research (AFSOR) award number FA9550-19-1-7010. This work made use of several open-source Python packages including pandas \citep{reback2020pandas}, Matplotlib \citep{Hunter:2007}, Scipy \citep{2020SciPy-NMeth}, Numpy \citep{harris2020array}, Cartopy \citep{Cartopy}, Astropy \citep{astropy:2013, astropy:2018}, and Sunpy \citep{sunpy_community2020}. The VLF data used in this statistical study are available at \url{data.rosseobservatory.ie/}, and the GOES/XRS data is available at \url{www.ngdc.noaa.gov/stp/satellite/goes-r.html} and were queried and downloaded using Sunpy. Our current VLF data from both observatories at Birr and Dunsink is now available at \url{vlf.ap.dias.ie/data/}. 
 The authors would like to thank the anonymous reviewer for the positive feedback and helpful comments, which improved this manuscript.
\end{acks}

{\footnotesize\paragraph*{Disclosure of Potential Conflicts of Interest}
The authors declare that they have no conflicts of interest.
}

%%%%%%%%%%%%%%%%%%%%%%%%%%%%%%%%%%%%%%%%%%%%%%%%%%%%%%%%%%%%%%%%%%%%%%%%%%%
%\appendix   

     % format of references provided by the journal (.bst)
\bibliographystyle{spr-mp-sola}
     % name your Bibtex file containing your references (.bib)
\bibliography{sola_vlf_paper}  

     % Checking: look if the file containing the ``\bibitem'' exits
     %           so check if the .bbl file exist (bibTeX compilation)
% \IfFileExists{\jobname.bbl}{} {\typeout{}
% \typeout{****************************************************}
% \typeout{****************************************************}
% \typeout{** Please run "bibtex \jobname" to obtain} \typeout{**
% the bibliography and then re-run LaTeX} \typeout{** twice to fix
% the references !}
% \typeout{****************************************************}
% \typeout{****************************************************}
% \typeout{}}

\end{article} 

\end{document}